\documentclass[article]{revtex4}
\usepackage{graphicx}% Include figure files

\begin{document}
%\title{Ultrafast acousto-magneto-plasmonics}
%\title{Plasmonics in hybrid multilayer structures: merging plasmonics, magnetism, acoustics and ultrafast optics}
\title{The marriage of plasmonics, magnetism, acoustics and ultrafast optics}

%% Notice placement of commas and superscripts and use of &
%% in the author list

\author{Vasily V. Temnov}

\affiliation{Institut des Mol\'ecules et Mat\'eriaux du Mans, UMR
CNRS 6283, Universit\'e du Maine, 72085 Le Mans cedex, France}

\date{\today}

\maketitle \textbf{Surface plasmon polaritons (SPP or SP) are
electromagnetic waves propagating along metal dielectric
interfaces and existing over a wide range of frequencies. They
have become popular because of their sub-wavelength confinement
and the possibility to perform ultrasensitive optical
measurements. Driven by tremendous progress in nanofabrication
techniques and ultrafast laser technologies the applications of SP
nanooptics extend beyond the border of nanoplasmonics. Here we
review how using novel hybrid multilayer structures combining
different functionalities allows to develop active plasmonic
devices and new metrologies. Magneto-plasmonics,
acousto-plasmonics and generation of high-energy photoelectrons
using ultrashort SP pulses represent a few examples how the
combination of ideas developed in the individual subfields can be
used to generate new knowledge suggesting plenty of exciting
applications in nanophotonics.}

\begin{figure}
\includegraphics[width=10cm]{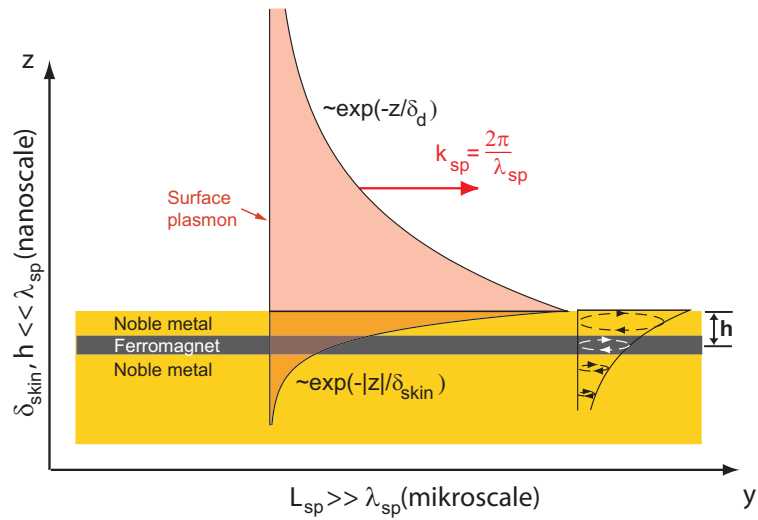} \caption{ {\bf
Surface plasmons in metal/ferromagnet/metal multilayer
structures.} A few nanometers thin ferromagnetic layer integrated
in the metal within SP skin depth $\delta_{\rm skin}$ almost does
not disturb the spatial distribution of SP intensity inside the
metal and can be used to control SP propagation via
magneto-optical effect in cobalt. Surface plasmon with frequency
$\hbar\omega=1.55$~eV (corresponding to $\lambda=800$~nm optical
wavelength in free space) at gold/air interface is characterized
by SP wavelength $\lambda_{\rm sp}=794$~nm, propagation distance
$L_{\rm sp}=45~\mu$m, skin depth $\delta_{\rm skin}=13$~nm and
decay length in air $\delta_d=307$~nm. Dashed elliptical contours
show trajectories of electrons moving in SP electric field.}
\end{figure}

Investigation of plasmonic devices beyond the diffraction limit
\cite{Gramotnev2010Nphot4} represents a topical research direction
in active plasmonics with major progress reported on the
development of SP-based amplifiers and lasers
\cite{Berini2012Nphot6}. After the original theoretical proposal
by Krasavin and Zheludev to use structural thermally induced phase
transitions in gallium-coated plasmonic waveguides
\cite{Krasavin2004APL84} many other ways to actively control SP
propagation have been realized in the experiment including
electro-optic effect in dielectric overlayers
\cite{Dicken08Nanolett}, optical excitation of semiconductor
quantum dots \cite{Pacifici07NPhot1,Fedutik07PRL99} or
magnetization switching in magneto-plasmonic materials
\cite{Temnov10NPhoton4}. Whereas nanoplasmonics with quantum dots
has been extensively discussed elsewhere
\cite{Berini2012Nphot6,QDPlasmonics} here we review the progress
in magneto-plasmonics, ultrafast acousto-plasmonic and
magneto-acoustic interactions. This development is largely driven
by continuously improving nanofabrication techniques and borrowing
the ideas from other well-established research directions. Despite
of being very different in nature these studies follow the same
idea: to explore various light-matter interactions in metallic
nanostructures at the true nanoscale, within the tiny skin depth
of light at visible and near IR frequencies. Development of hybrid
ultrafast nanophotonic devices for future telecommunication and
data recording technologies represents the final goal of these
research activities.

The basic knowledge of the spatial distribution of electromagnetic
field at interfaces supporting propagating SP waves appears to be
inevitably necessary to understand the underlying physics. A
typical hybrid noble metal/ferromagnet/noble metal multilayer
structure, to be discussed within the framework of
magneto-plasmonics, is sketched in Fig.~1. The strong dielectric
contrast at metal-air interface characterized by a large negative
real part of metal dielectric function determines the spatial
distribution of exponentially decaying SP electric field. Due to a
to a much weaker dielectric contrast between the adjacent metals
the spatial distribution of SP intensity in a
metal/ferromagnet/metal trilayer exhibits only minor deviations as
compared to SP at single metal-air interface ($|E_y|^2$ is shown
in Fig.~1) \cite{Temnov10NPhoton4}. In order to be able to
efficiently interact with surface plasmons, the ferromagnetic
layer must have substantial overlap with the electric field of
surface plasmon. SP intensity penetration (skin) depth
$\delta_{\rm skin}$ inside the metal is of the order of 10~nm in
the visible and near infrared frequency range dictating the
ultimate nano-scale for hybrid noble metal/ferromagnet multilayers
characterized by $h\sim\delta_{\rm skin}\ll\lambda_{\rm
sp}\simeq\lambda$. SP decay length $\delta_d$ on the dielectric
side is a fraction of optical wavelength $\lambda$ and is thus
significantly larger as compared to the skin depth.

Strongly absorbing ferromagnetic metal layer introduces
substantial losses - both due to large intrinsic absorption and
possibly due to increased interface roughness. However, SP
propagation distance in such hybrid structures usually remains
large compared to optical wavelength: $L_{\rm sp}\gg\lambda_{\rm
sp}$ (for example $L_{\rm sp}=10~\mu$m for a Au/Co/Au structure
with h=8~nm and 6~nm thickness of the cobalt layer
\cite{Temnov10NPhoton4}). This large propagation distance allows
to design devices with plasmonic in- and out-couplers separated by
large distance $D\sim L_{\rm sp}$ of the order of at least a few
microns and perform optical measurements with conventional optical
microscopy techniques. Small external modulation of SP wave vector
$\delta k_{\rm sp}$ or/and decay length $\delta L_{\rm sp}$ is
accumulated over a large SP propagation distance and can be easily
measured by surface plasmon interferometry or, to some extent, in
Kretschmann configuration (see Box1).

\subsection{Magneto-plasmonics}

\begin{figure}
\includegraphics[width=12cm]{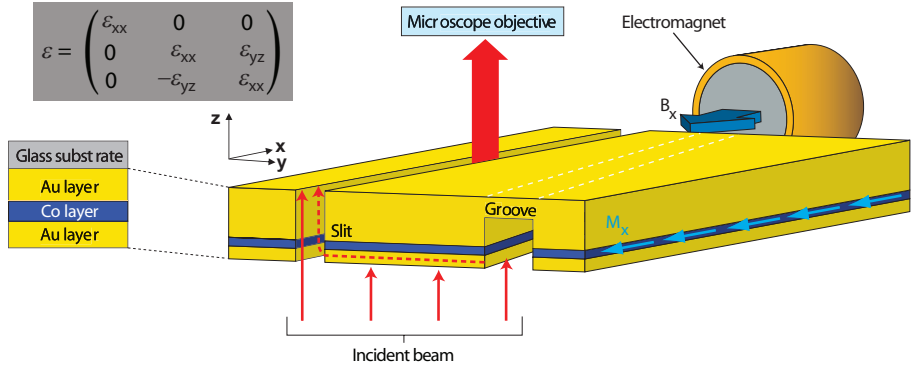} \caption{{\bf
Active magneto-plasmonics.} The magnitude of SP wavevector $k_{\rm
sp}(M_x)$ in a magneto-plasmonic Au/Co/Au trilayer changes as the
in-plane magnetization $M_x$ in cobalt is periodically switched by
a weak (20~mT) external magnetic field of an electromagnet, thus
shifting the optical phase in the plasmonic interference arm by
$\delta k_{\rm mp}D$=0.02~rad. Figure adopted with permission from
ref.~\cite{Fan10NP_news}.}
\end{figure}

The idea to control the optical properties of SP using magnetic
effects started with theoretical paper by Chiu and Quinn
\cite{Chiu72NuoveoCimento10}, who investigated the change in SP
dispersion relation at an interface between vacuum and
free-carrier metal induced by an external magnetic field. The
Lorentz force acting on the the electrons moving along the closed
elliptical trajectories (dashed contours in Fig.~1) in the
electric field of SP was shown to change the magnitude of SP wave
vector. The largest change was predicted for in-plane magnetic
field perpendicular to SP wave vector (in x-direction). Exploiting
the high sensitivity of Kretschmann configuration Haefner and
co-workers \cite{Haefner94PSS185} observed this tiny effect in
plasmonic free-carrier like metals Cu, Ag, Au and Al at relatively
low external magnetic field $B=200$ millitesla (mT). A good
quantitative agreement was obtained for Ag and Au using simple
free-electron (Drude) theory for the non-diagonal components of
the linear dielectric permittivity tensor
\begin{equation}
\label{NondiagElement} \epsilon^{Drude}_{\rm
yz}=i\frac{\omega_{\rm c}\omega_{\rm
p}^2}{\omega[(\omega+i/\tau_{Dr})^2-\omega_{\rm c}^2]}\,,
\end{equation}
where $\omega_{\rm c}=eB_x/m_{eff}$ is a cyclotron frequency ($e$
- electron charge, $m_{eff}$ - electron effective mass). Whereas
expression (\ref{NondiagElement}) provided accurate results for Ag
and Au, only order of magnitude agreement for Al and Cu was
obtained. It was explained by possible contributions of interband
transitions, which are disregarded in the Drude model. The
magnitude of magneto-optic effect in diamagnetic noble metals
$\epsilon_{yz}/\epsilon_{xx}\sim 10^{-7}B_x$[mT] was found too
small ruling out practical applications requiring reasonably low
external magnetic fields.

Much larger magneto-optical effects with
$\epsilon_{yz}/\epsilon_{xx}\sim 10^{-2}$ were found in
ferromagnets . Modern energy-band theory calculations suggest that
the magneto-optical properties in Fe, Ni and Co at visible
frequencies are dominated by interband transitions, with
$\epsilon_{yz}\propto M_x$ proportional to magnetization $M\sim
1$~Tesla in a ferromagnet \cite{OppeneerLPN}. Possible Drude-like
contributions caused by the very same Lorentz force in a large
built-in magnetic field of a ferromagnet $B_{eff}\propto M$
\cite{Krinchik64JAP35} are shown to contribute to the non-diagonal
elements of dielectric permittivity only in the near-infrared
frequency range (for photon energies below 1 ~eV in
Fe)\cite{OppeneerLPN}. Since both interband and intraband (Drude)
contributions scale linearly with $M$, optical measurements of the
non-diagonal components are widely used to study magnetic
properties of materials, including at ultrafast time scales.

Experiments with multilayer structures consisting of a thin
ferromagnetic film sandwiched between two layers of noble metal in
Kretschmann configuration demonstrated further increase of
magneto-optical effects due to excitation of SP
\cite{Hermann2001PRB64}. Systematic studies of Au/Co/Au trilayers
with different geometric parameters showed that structures with a
6~nm thin cobalt layer gave maximum enhancement of
magneto-plasmonic signal in Kretschmann configuration
\cite{Gonzalez-DiazPRB}. Another advantage of using thin
ferromagnetic layers is that a small external magnetic field of
the order of a few millitesla (mT) is sufficient to switch the
in-plane magnetization. The magnitude of the switching field in
single crystal samples drops below $1$~mT for thin epitaxial iron
films, one order of magnitude smaller than in polycrystalline
samples, opening the door for practical applications
\cite{FerreiroVila2011PRB83}.

Combined with SP interferometry these studies allowed to build a
prototype of magneto-plasmonic device consisting of a tilted
slit-groove microinterferometer milled by a focussed ion beam in
the magneto-plasmonic Au/Co/Au (5~nm/6~nm/191~nm) multilayer
structure \cite{Temnov10NPhoton4}. The external magnetic field of
an electromagnet switched the magnetization of the cobalt layer
and thus changed the magnitude of SP wave vector. The relatively
small magneto-plasmonic modulation of SP wavevector $\delta k_{\rm
mp}/k_{\rm sp}\sim 10^{-4}$ was accumulated over a large
propagation distance $D=22~\mu$m and produced a substantial shift
of plasmonic interference fringes of 0.02 radian between two
opposite magnetization directions. This phase shift could be
further increased by covering the device with a thin layer of a
high-index dielectric material \cite{MartinBecerra2010APL97},
pointing towards future application as a magneto-plasmonic switch.

The magneto-plasmonic modulation of SP wave vector is ultimately
sensitive to SP intensity at the location of a thin ferromagnetic
layer. Therefore it can be used to measure the electric field
distribution inside the metal with nanometer spatial resolution
determined by the thickness of the magnetic (probe) layer. By
varying the position $h$ of a 6~nm thin cobalt layer the 13~nm
skin depth of surface plasmon in gold was directly measured
\cite{Temnov10NPhoton4}.

A slightly different version of magneto-plasmonic measurements was
implemented in periodically perforated metal films covered with a
layer of a ferromagnetic dielectric bismuth iron garnet
\cite{Belotelov2011NNanotechn6}. Here, the ability to selectively
modify SP wavevector at metal/magnetic dielectric interface not
only confirmed the plasmonic origin of extraordinary high
transmission \cite{Ebbesen98Nature391} but also demonstrated a
giant magneto-optical Kerr effect in transmission mode. In another
experiment the surface-sensitive non-linear optical technique of
magnetization-induced second harmonic generation revealed the
direction of magnetization in nickel nanostructures, an effect
interesting for magneto-plasmonic sensing applications
\cite{Valev2011ASCNano5}. The combination of magneto-plasmonic
detection in a periodically oscillating magnetic field and
Kretschmann geometry was shown to increase the detection
sensitivity for biosensing applications \cite{Regatos11OE19} as
compared to conventional SPR sensors \cite{Homola99SAB54}.

One of the most intriguing questions concerns the maximum possible
speed of all-optical modulation in nanophotonic devices and
limiting time of magnetization reversal (switching) in
magneto-plasmonic and, more generally, magneto-optical devices,
which can be potentially used in new ultrafast telecomminication
and data recording technologies. These questions have been tackled
in various experiments combining magnetism, plasmonics with
ultrafast optical measurements.

\subsection{Ultrafast opto-plasmonic modulation}

\begin{figure}
\includegraphics[width=10cm]{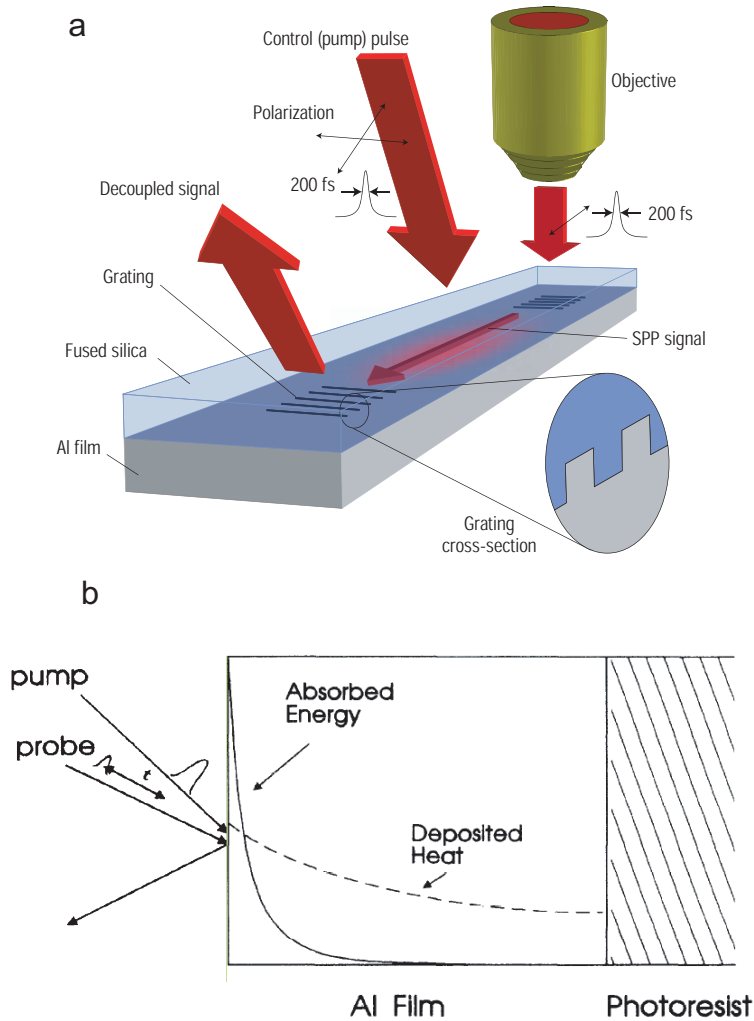} \caption{{\bf Ultrafast
plasmonic modulation in fs-laser excited aluminum films.} {\bf a,}
The transmission of an ultrashort SP pulse propagating at
Al/silica interface is perturbed by focussing an intense
femtosecond control (pump) pulse between the gratings, which serve
for SP in- and out-coupling. {\bf b,} Efficient electron diffusion
in plasmonic metals leads to much larger heat penetration depth as
compared to the skin depth, where the energy of an intense
femtosecond pump pulse is absorbed, the effect known from
conventional time-resolved pump-probe reflectivity measurements.
Figures reproduced with permission from: {\bf a,}
ref.~\cite{MacDonald08NPhot3};{\bf b,} ref.~\cite{Tas94PRB49}.}
\end{figure}

A modern version of femtosecond time-resolved measurements in a
nanophotonic device was reported recently by MacDonald and
co-workers \cite{MacDonald08NPhot3}, who have designed an
ultrafast plasmonic light modulator (Fig.~3a). A pair of identical
one-dimensional gratings was etched in silica and covered with
aluminum film by thermal evaporation. A femtosecond laser pulse
illuminated the first grating launching an ultrashort SP pulse,
which propagated at aluminum-silica interface and was diffracted
into free-space radiation by the second grating. An intense
ultrashort control pump pulse was focused between the gratings and
triggered diverse nonlinear optical phenomena in aluminum and,
therefore, changed the optical absorption for ultrashort SP
pulses. By varying the delay between the pump (control) and probe
(signal) pulses the researchers demonstrated ultrafast plasmonic
modulation of the decoupled signal on a sub-picosecond time scale
opening the door to all-optical modulation rates in the THz
frequency range. The origin of ultrafast sub-picosecond modulation
in this device was attributed to the pump-induced third-order
optical nonlinearity in the vicinity of interband transition in
aluminum at $\lambda=800$~nm \cite{Samson2009JOA11} and possible
contribution from transient grating effect originating from the
interference between pump and SP pulses. A clean observation of
transient grating effect with surface plasmons was reported later
in a different experiment on planar gold-air interface
\cite{Rotenberg2010PRL105}.

However, the device in Fig.~3a also showed much slower dynamics
evident on the time-scale of tens of picoseconds. To explain these
a variety of physical phenomena following the interaction of a
femtosecond laser pulses with metal needs to be discussed in more
detail \cite{DelFatti00PRB61}. An ultrashort laser pulse is
absorbed by free carrier absorption creating a short-living
distribution of non-equilibrium electrons. Electrons thermalize
with each other through electron-electron collisions within at
most a few hundreds of femtoseconds, with the longest electronic
thermalization time of $\sim$500~fs reported for gold
\cite{DelFatti00PRB61}. After thermal equilibrium between the
electrons is established, which can be adequately described by a
high electron temperature, the transient state is still highly
non-equilibrium because the electron temperature is still much
higher than lattice temperature. Due to electron-phonon collisions
hot electrons cool down and the lattice warms up until the common
equilibrium temperature is established within a few picoseconds.
This sequence of events in time domain is accompanied by diffusion
of hot electrons in space, out of the skin depth where they
originally absorbed laser energy. Later on, thermal expansion
leads to the generation of coherent acoustic pulses
\cite{Thomsen86PRB34} followed by cooling down of the lattice due
to thermal diffusion on the time scale exceeding tens of
picoseconds.

An elegant experiment exploiting thermo-acoustic generation as a
probe of electronic heat diffusion depth in noble metals was
reported by Tas and Maris \cite{Tas94PRB49}. For sufficiently thin
samples, a situation sketched in Fig.~3b, the heat diffusion by
hot electrons generated by the pump pulse leads to the nearly
homogeneous heating of an aluminum film. It resulted into the
generation of the acoustic pulse at aluminum-photoresist interface
caused by thermal expansion of aluminum. After propagation through
the film at the speed of sound $c_s$ the acoustic strain pulse
arrived at aluminum-air interface where it was detected by a
time-delayed probe pulse via strain-induced transient reflectivity
changes, a standard technique in picosecond ultrasonics
\cite{Thomsen86PRB34}.

\subsection{Ultrafast acousto-plasmonics}

\begin{figure}
\includegraphics[width=10cm]{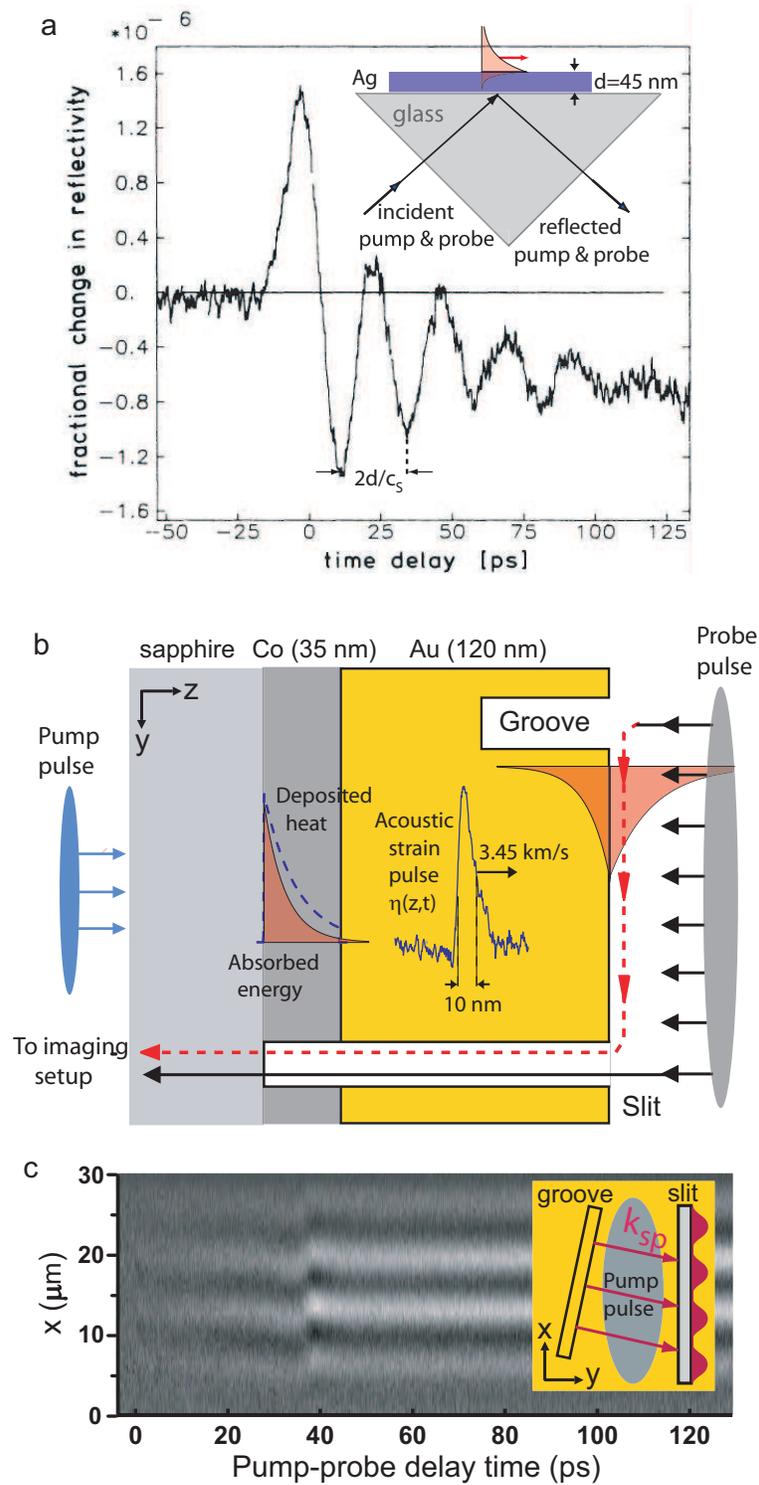} \caption{{\bf
Ultrafast acousto-plasmonics.} {\bf a,} Plasmonic pump-probe
experiments in Kretschmann geometry allow for ultrasensitive
detection of coherent acoustic oscillations induced by thermal
expansion of a 45~nm thin silver layer homogeneously heated by an
ultrashort laser pulse. {\bf b,} In hybrid acousto-plasmonic
structures the thermal expansion of cobalt transducer excited by a
femtosecond pump pulse at time zero launches an ultrashort
acoustic pulse $\eta(z,t)$ propagating through gold at the speed
of sound $c_s=3.45$~km/s. {\bf c,} After approximately 38~ps the
acoustic reflection from gold-interface changes the wavevector of
a time-delayed ultrashort SP probe pulse generating the
acousto-plasmonic pump-probe interferogram in a tilted slit-groove
arrangement. The 10~nm wide acoustic strain pulse in Fig.~4b has a
duration of 3~ps and is reconstructed from transient phase shift
in Fig.~4c. Measurements are performed by time-resolved SP
interferometry \cite{SPInterferometry09OE}. Figure {\bf a,}
adopted with permission from ref.~\cite{vanExter88PRL60}.}
\end{figure}

It was demonstrated by van Exter and Lagendijk back in 1988 that
SP excitation in Kretschmann geometry can greatly enhance the
magnitude of pump-probe signals as compared to conventional
reflectivity measurements \cite{vanExter88PRL60}. A thin silver
layer was excited through the glass prism by a picosecond pump
pulse resulting into the homogenous heating through film depth
$d=45$~nm. The dynamics of thermal expansion governed by acoustic
generation, propagation and reflection at silver-glass and
silver-air interfaces caused the periodic expansion and
contraction of silver layer at a frequency of 42~GHz=c$_s/(2d)$
determined by the acoustic round trip through the layer. Tiny
acoustic modulation in lattice density changed the wavevector of
time-delayed SP probe pulses resulting into drastic change of
probe reflectivity in Kretschmann configuration. From the decay
time of thermo-acoustic oscillations, which was mainly due to the
finite (52$\%$) acoustic reflection at silver-glass interface, the
researchers were able to conclude that the intrinsic life time of
longitudinal phonons in silver at 42~GHz largely exceeded 100~ps.
Nowadays measurements of the mean free path (or lifetime) of
high-frequency phonons in solids are found to be extremely
important to understand thermal properties of the materials at the
nanoscale driven by the interplay of ballistic and diffusive heat
transport \cite{Siemens2010NMat9}. Similar fs-time-resolved
experiments in thin gold and copper films
\cite{Wang2007OL32,Wang2007PRB75} revealed the contribution of
short-living non-equilibrium electrons in the build up of
mechanical stress driving the initial stage of thermo-acoustic
expansion, in agreement with analogous measurements in plasmonic
nanoparticles \cite{Perner2000PRL85}.

Despite of the appealing simplicity and high sensitivity of
plasmonic pump-probe experiments in Kretschmann configuration
their quantitative yield appeared to be rather limited. Very
accurate time- and angular-dependent reflectivity measurements
were necessary to reconstitute the dynamics of surface dielectric
function \cite{Groeneveld90PRL64}. Quantitative measurements can
be performed by plasmonic pump-probe interferometry with
femtosecond time resolution \cite{SPInterferometry09OE}. Figure 4b
shows the experimental configuration of this technique adopted to
acousto-plasmonic studies. In contrast to most previous
measurements performed with a single polycrystalline metal layer
on a dielectric substrate we used a Au/Co/sapphire structure with
well-defined crystallographic (111) orientation of gold
\cite{TemnovAcoustoPlasmonics}. A 35~nm thick cobalt layer excited
by an intense ultrashort laser pulse through sapphire substrate
serves as an efficient and ultrafast opto-acoustic transducer. Due
to much faster electron-phonon relaxation in ferromagnetic
materials of the order of 0.2-0.3~ps \cite{Koopmans2010NMat9} and
much shorter electronic mean free path $l_e\sim 1$~nm
\cite{Getzlaff93SSC87} the hot electron diffusion is much less
efficient than in noble metals characterized by $l_e\sim 40$~nm
\cite{Crowell62PR127}. Therefore the heat penetration depth in
ferromagnetic metals only slightly (typically by $\sim50\%$)
exceeds the skin depth for pump light \cite{Saito03PRB67}. Thermal
expansion of fs-laser-heated cobalt layer generates an ultrashort
acoustic strain pulse in both directions, into the gold layer and
into the sapphire substrate. Very good acoustic impedance matching
between three layers suppresses the acoustic reflection at
gold/cobalt and cobalt/sapphire interfaces (both are about $10\%$)
and the initial shape of a unipolar acoustic pulse follows the
spatial profile of deposited heat in cobalt. Keeping in mind
interesting nonlinear dynamics of ultrashort acoustic pulses
observed in sapphire \cite{Hao01PRB64,vanCapel10PRB81} here we
focus on the acousto-plasmonic effect in gold. The compressional
acoustic pulse in gold $\eta(z,t)=(n_i(z,t)-n_i^0)/n_i^0$ creates
a layer of higher ion density $n_i(z,t)>n_i^0$ and moving slowly
at sound velocity $c_s$=3.45~km/s. Due to the strong acoustic
anisotropy in crystals the speed of sound in (111) direction is
significantly higher than the value of 3.24~km/s in
polycrystalline gold, averaged over different grain orientations.
Since the stationary charge separation between the electrons and
the ions in a metal may occur only within a tiny Debye radius
$r_{\rm Debye}\sim 10^{-3}$~nm, the spatial profile of electron
density $n_e(z,t)$ exactly follows the ionic one:
$n_e(z,t)=n_i(z,t)$. At 1.55~eV photon energy ($\lambda = 800$~nm)
of probe pulses the dielectric function
$\epsilon_m=\epsilon^{'}+i\epsilon^{''}=-24.8+1.5i$ in gold is
dominated by free-carrier contribution with $\epsilon^{'}\simeq
-\omega^2_{\rm p}/\omega^2\propto -n_e$. An ultrashort acoustic
strain pulse creates a time-dependent spatial profile of the
dielectric function $\epsilon^{'}(z,t)= \epsilon^{'}(1+\eta(z,t))$
inside the metal, which modulates SP wavevector
\cite{SPInterferometry09OE}
\begin{equation}
\label{AcPlasmEquation} \delta k_{\rm
sp}(t)=\frac{k_0}{2|\epsilon_m|\delta_{\rm skin}}
\int\eta(z,t){\rm exp}(-|z|/\delta_{\rm skin}){\rm sgn}(z-c_st)dz
\,,
\end{equation}
when the strain pulse arrives within SP skin depth. Here the ${\rm
sgn}(z-c_st)$ function describes the transformation of the
incident compressive pulse into the tensile one during the
acoustic reflection at metal-air interface. Experimental
acousto-plasmonic pump-probe interferogram in Fig.~4c demonstrates
a pronounced transient shift of plasmonic fringes governing the
acoustic reflection at gold-air interface. Application of standard
interferometric analysis techniques and using expression
(\ref{AcPlasmEquation}) allows to reconstruct the acoustic pulse
shape without using any fit parameters, see the acoustic pulse in
Fig.~4b. Acoustic pulse width of 10~nm (FWHM) corresponds to pulse
duration of 3~ps. The temporal profile of the acoustic pulse is
measured with 600~fs time resolution, which is limited by
nanometer surface roughness ($\sim 2$~nm RMS) implying random
distribution of acoustic arrival times at gold-air interface. The
peak strain of presented acoustic pulse is $\eta_{max}=2.5\times
10^{-3}$ although much higher strain amplitudes up to 0.01
(corresponding to uniaxial stress of $\sim 2$~GPa) and leading to
the non-linear acoustic propagation effects in gold were observed
in the same sample \cite{TemnovAcoustoPlasmonics}. Application of
1$\%$-strain in gold shifts the plasmonic fringes by
$\sim$0.01~radian, similar to the magneto-plasmonic switch in
Fig.~2. The time scale of ultrafast acousto-plasmonic modulation
is limited by the acoustic travel time through the skin depth
$\tau_{\rm skin}=\delta_{\rm skin}/c_s=3.8$~ps (in gold), although
much shorter pulses can be detected.

Being ultimately sensitive to free-carrier density the plasmonic
detection represents a quantitative technique which is
complimentary to the acoustic characterization via pump-probe
reflectivity measurements \cite{Thomsen86PRB34} and transient
surface deformations with common-path interferometry
\cite{vanCapel10PRB81}. Indeed, particularly large pump-probe
signals observed in metals in the vicinity of interband
transitions \cite{Thomsen86PRB34} show drastic change in shape due
to strong and often unknown wavelength-dependence of photoelastic
coefficients $d\sqrt{\epsilon_m}/d\eta$ \cite{Devos01PRL86}. An
encouraging way to obtain photo-elastic coefficients in noble
metals is provided by recent studies of tiny plasmonic
nanoparticles. Large surface tension in nanoparticles with
nanometer radius leads to their contraction and changes the
electron density inducing the blue shift in plasma frequency
\cite{Cai2001EPJD13}. However, if the optical wavelength falls
within the region of interband transitions, lattice contraction
also changes the dielectric function of core electrons according
to simple Clausius-Mosotti formula inducing the red-shift of SP
resonance frequency \cite{Lerme2001EPJD17}. Although this theory
has not been rigorously compared with experimental data in
picosecond ultrasonics, the competition of intraband
(free-carrier) and interband contributions provides a reasonable
explanation for disappearance of acousto-optic pump-probe signals
in Cu and Al in the vicinity of interband transitions (at
$\lambda$=575~nm and 850~nm, respectively) \cite{Devos03PRB68}. In
case the reliable tabulated values for photoelastic coefficients
in noble metals become available simple pump-probe reflectivity
measurements may be able to provide quantitative characterization
of ultrashort acoustic strain pulses as well. Extraction of
acoustic pulse shapes from much more complicated time-resolved
interferometry measurements also depends on photo-elastic
coefficients, although under some specific conditions the measured
optical phase
$\delta\phi(t)=\delta\phi_{metal}(t)+\delta\phi_{surface}(t)$ is
dominated by the second well-defined term
$\delta\phi_{surface}(t)=(4\pi/\lambda)\delta z(t)$ caused by
transient surface displacement $\delta z(t)$
\cite{vanCapel10PRB81,Temnov06JOSAB23_1954}.

SP propagation along the surface is usually not affected by small
transient surface deformation $\delta z(y,t)$. However, spatially
periodic modulation $\delta z(y)=\delta\sin{(2\pi y/\Lambda_{\rm
G}})$ along SP propagation direction $y$ may serve as a grating
for phase-matched in- and out-coupling of SP into free space
radiation. Periodic grating at gold-air interface with amplitude
$\delta\sim 1$~nm and $\Lambda_{\rm G}=7.6~\mu$m induced by
surface acoustic wave at 480~MHz frequency was shown to convert
near-infrared light into SP with $\sim 10^{-4}$ efficiency
\cite{Ruppert2010PRB82}.

\subsection{High-energy photoelectron generation with femtosecond
SP pulses}

\begin{figure}
\includegraphics[width=10cm]{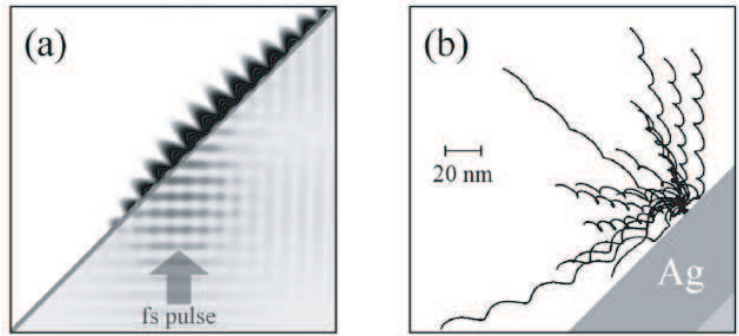} \caption{{\bf Generation
and acceleration of high-energy photoelectrons with fs-laser
pulses in Kretschmann geometry.} {\bf a,} The intensity of
incident fa-laser pulse is strongly enhanced as it is converted
into an ultrashort SP in a 50~nm thin silver layer in Kretschmann
configuration. {\bf b,} Calculated trajectories of photoelectrons
accelerated away from the surface in the field of intense
ultrashort SP pulse. Figures reproduced with permission from
ref.~\cite{Irvine2004PPL93}.}
\end{figure}

Another interesting application in ultrafast plasmonics reveals
the possibility to generate high-energy photoelectrons at
metal-vacuum interface using femtosecond SP pulses. This
phenomenon was observed both with an amplified high-power 250~kHz
laser system \cite{Zawadzka2000NIMPRA445} and a non-amplified
Ti:sapphire oscillator operating at high repetition rate of 80~MHz
\cite{Irvine2004PPL93}. In these experiments femtosecond laser
pulses were used to illuminate a 50~nm thin silver layer in
Kretschmann geometry (Fig.~5a). The combination of ultrashort
pulse duration, intrinsic field enhancement in Kretschmann
geometry and sub-wavelength confinement of SP intensity
exponentially decaying within $\delta_d=240$~nm allowed to reach
extremely high gradient of electric field of $8\times
10^{13}$~V/cm$^2$ in the direction normal to the surface
\cite{Irvine2004PPL93}. Figure 5b shows typical calculated traces
of photoelectrons, which were born at random times and with random
initial velocities in three-photon-photoemission process and
accelerated away from the surface on quivering trajectories
gaining the kinetic energy up to 0.4~keV \cite{Irvine2004PPL93}.
Most recent experiments with amplified phase-stabilized laser
pulses allowed to reach electron kinetic energies up to 100~keV
and also revealed the importance of nanoscale surface roughness
affecting the photoemission process of individual electrons
\cite{Racz2011APL98}.

Photoemission and acceleration of photoelectrons in SP electric
field generates an ultrashort pulse of electric current in the
direction normal to the surface. SP-mediated THz radiation emitted
by this current was observed very recently in periodic gratings on
gold and silver surfaces \cite{Welsh2009OE17}. The dependence of
THz intensity on the angle of incidence and polarization of
intense femtosecond pump pulses clearly demonstrated the
dominating role of SP in THz generation. These experiments explain
the diversity of published experimental data (see references in
\cite{Welsh2009OE17}) and undermount the hypothesis about the role
of non-equilibrium ballistic electron transport, which was
suggested to explain the non-local character of THz generation in
thin gold films \cite{Kadlec2005OL30}.

\subsection{Ultrafast laser-induced magnetization dynamics}
Besides the complex spatio-temporal dynamics of electrons and
phonons in noble metals the femtosecond optical measurements in
ferromagnets are even more important. Using femtosecond
Kerr-rotation measurements Beaurepaire and co-workers demonstrated
\cite{Beaurepaire96PRL76} that fs-laser induced heating of
electrons resulted into the simultaneous drop in magnetization
(ultrafast demagnetization) in a ferromagnetic nickel. Later on it
has been shown that the demagnetization dynamics occured within
electron-electron thermalization time of about 50~fs after which
the spins followed the dynamics of electronic temperature
\cite{Guidoni02PRL89}, as confirmed by other measurements
\cite{Guedde99PRB59}. For large density of laser excitation
(fluence) of the order of a few mJ/cm$^2$ the demagnetization
dynamics became even more complex and could be described by
different relaxation times for electrons and spins, both in
sub-100~fs time range \cite{Bigot09NPhys5}.

Ultrafast decay of magnetization in Ni was contrasted by
relatively slow demagnetization in Gd on a time-scale of about
100~ps \cite{Vaterlaus91PRL67}. Very recently the paradoxical
diversity of magnetization dynamics observed in many magnetic
materials was explained by different contributions to
magnetization via delocalized electronic states (3d4sp states
dominating in Ni) and bound states (4f states in Gd)
\cite{Koopmans2010NMat9}. The observation of small ultrafast
component in magneto-optical Kerr measurements in Gd caused by
heating of 5d6sp electrons confirmed the universal character of
ultrafast demagnetization pathway by delocalized electrons.

The dynamics of magnetization recovery is governed by its
precessional motion and can be adequately described by the
phenomenological Landau-Lifshitz-Gilbert (LLG) equations in which
the length of magnetization vector is not conserved
\cite{Vomir05PRL94}: the typical magnetization dynamics represent
a damped precession with a period of several tens of picoseconds
and a decay time of a few hundreds of picoseconds.
Magnetocrystalline anisotropy in cobalt is found to significantly
alter the trajectories of magnetization vector, an effect
important in the context of studying switching dynamics
\cite{Bigot05ChemPhys318}.

Numerous studies of fs-laser induced magnetization dynamics in
various ferromagnetic and antiferromagnetic materials lead to the
pioneering observation that a single intense circularly polarized
fs-laser pulse can deterministically switch the magnetization in a
ferrimagnetic GdFeCo alloy, with the magnetization direction
determined by the helicity (right- or left-handed) of circularly
polarized light \cite{Stanciu2007PRL99,Vahaplar2009PRL103}. The
switching pathway could be adequately described by LLG equations
for magnetization dynamics as well. Very recently this experiment
was reproduced with much longer picosecond pump pulses confirming
the precessional nature of magnetization switching
\cite{Steil2011PRB84}. However, the physical nature of the
long-living helicity reservoir for angular momentum in the
material was called into question for a following reason: the
angular momentum cannot be stored long enough in short-living
electronic excitations as comes out to be necessary to explain
switching with picosecond laser pulses. Irrespective of the pulse
duration switching occurred only when a significant amount of heat
(incident laser fluence of 6 mJ/cm$^2$) was deposited in the
material. At even higher fluences the deterministic thermal
magnetization switching was observed with linearly polarized light
\cite{Ostler2012Ncomm3} suggesting that all-optical
helicity-dependent reversal exists only within a relatively narrow
range of fluences (within $\sim 10\%$ above the switching
threshold).

\subsection{Prospects: ultrafast acoustic magnetization switching}

For practical applications it would be highly desirable to develop
new ways of magnetization reversal not relying on thermal
mechanisms. The proof-of-principle non-thermal magnetization
reversal in Co/Pt multilayer structures was achieved with
ultrashort pulses of magnetic field $B(t)$ generated by
relativistic electron bunches \cite{Back98PRL81}, i.e. in a
large-scale facility experiment. Thinking further in terms of
miniaturized ultrafast hybrid devices combining different
functionalities one could imagine magnetization reversal in the
magneto-plasmonic switch by means of ultrashort acoustic pulses.

The possibility of using reasonably low external stress of the
order of 100~MPa as a source of an effective magnetic field
$B_{eff}(t)$ for magnetization reversal in nickel nanomagnets was
discussed recently \cite{Atulasimha2010APL97}. The first
experiments on acoustically induced magnetization dynamics in
ferromagnetic semiconductor GaMnAs at cryogenic temperature
revealed only minor perturbations of magnetization direction
\cite{Scherbakov10PRL105}. Similar room-temperature experiments in
polycrystalline nickel films demonstrated much larger acoustic
rotation of magnetization direction of about 4$^{\circ}$ out of
the equilibrium \cite{Kim_arXiv}.

It is likely that not only longitudinal (compresssional or
tensile) but also transversally polarized acoustic ${\it shear}$
pulses \cite{Pezeril09PRL102} can efficiently interact with
magnetization and may even contribute to the dynamics of
all-optical magnetization switching \cite{Stanciu2007PRL99}.
Femtosecond circularly polarized light pulses may act as an
ultrafast 'screwdriver': due to efficient electron-phonon coupling
the directed circular surface current driven by the rotating
electric field of a circularly polarized pump pulse should
generate a helically polarized pulse of acoustic shear phonons
acting as a source of long-living effective magnetic field
$B_{eff}(t)$ in the direction normal to the surface
\cite{delaFuente04JPCM16}. Whether or not this hypothesis, which
at least does not contradict the most recent experimental
observations \cite{Steil2011PRB84}, will be confirmed in the
experiment, the future of ultrafast magneto-elastic interactions
looks bright.

\subsection{BOX1: measurements with surface plasmons}

\begin{figure}
\includegraphics[width=12cm]{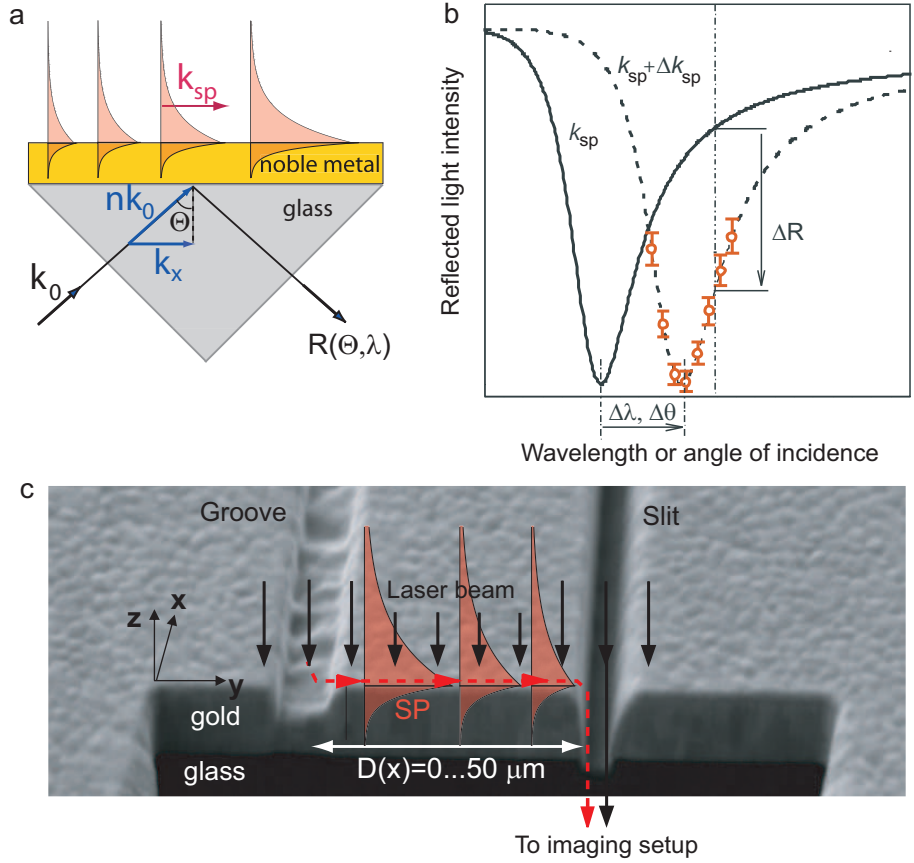} \caption{ {\bf
SP excitation and sensing.} {{\bf a,} Kretschmann geometry for SP
excitation in a thin film of a noble metal on a dielectric prism.
SP intensity grows upon propagation. {\bf b,} A small change of SP
wave vector $k_{\rm sp}$ induces a large change in the intensity
of reflected light $\Delta R$. {\bf c,} Scanning electron
microscopy image of a slit-groove pair milled by a focussed ion
beam in a 200~nm thin gold film on glass. A 200~nm wide and 100~nm
deep groove and a 100~nm wide slit are milled by the focussed ion
beam. SPs are exited at the groove, propagate towards the slit
where they interfere with directly transmitted light. SP intensity
decays upon propagation. Figures adopted with permission from:
{\bf b,} ref.~\cite{Piliarik72OC6}; {\bf c,}
ref.~\cite{SPInterferometry09OE}.}}
\end{figure}

When the electrons in bulk metal are displaced with respect to the
ions, the (tiny) charge separation induces an electric field,
which tends to restore the electrical neutrality. This relaxation
process is governed by oscillations of the electrons at plasma
frequency $\omega_{\rm p}=\sqrt\frac{n_ee^2}{\epsilon_0m_{eff}}$.
Since in most metals there is at least one free electron per ion,
the resulting electron density $n_e\sim 10^{22}$~cm$^{-3}$ is very
high and $\hbar\omega_{\rm p}\sim$10~eV. A simple Drude model for
the linear dielectric susceptibility of a metal
$\epsilon_m(\omega)=\epsilon_{xx}(\omega)=1-\omega_{\rm
p}^2/(\omega(\omega+i/\tau_{Dr}))$, where $\tau_{Dr}$ stands for
an effective electron scattering time, represents a reasonable
approximation for the optical properties of noble
free-carrier-like metals (Ag,Au,Cu,Al) in the spectral region free
from interband transitions. Their relatively low optical losses
due to the large electronic collision time $\tau_c\sim 10$~fs
favor applications in plasmonics, in contrast to high-loss
ferromagnetic metals (Fe,Ni,Co) characterized by $\tau_{Dr}\sim
1$~fs. Another very important parameter, the mean free path $l_e$
of electrons at a Fermi surface is also much larger in noble
metals.

Surface plasmon polaritons (SPP or SP) exist in a broad frequency
band obeying the inequality ${\rm Re}[\epsilon_m(\omega)]<-1$
($\omega<\omega_{\rm p}/\sqrt{2}$) and are characterized by
dispersion relation $k_{\rm
sp}(\omega)=k_0\sqrt\frac{\epsilon_m(\omega)}{\epsilon_m(\omega)+1}$,
where $k_0=\omega/c$ is the wavevector of light in vacuum and
$k_{\rm sp}$ is SP wave vector. Since $k_{\rm sp}>k_0$ for all
frequencies SP cannot be excited by plane electromagnetic waves
impinging on a metal-air interface. An elegant and most widely
used way to excite surface plasmons is provided by the so-called
Kretschmann geometry \cite{Kretschmann68ZNA23}(see Fig.~6a), where
a thin metal film deposited on top of a dielectric prism is
illuminated by a collimated light beam through the prism with
refractive index $n>1$. For a particular combination of optical
wavelength and angle of incidence $\Theta_0$ the in-plane
component of the wave vector of p-polarized incident light
$k_x(\Theta)=nk_0\sin\Theta$ becomes equal to SP wave vector,
$k_{\rm sp}(\Theta_0)=k_x(\Theta_0)$, leading to phase-matched
excitation of SP at metal-air interface. The angular dependence of
reflectivity $R(\theta,\lambda)$ shows a very sharp dip at
$\theta=\theta_0$ with a width of about 1 degree, see Fig.~6b.
Thanks to very high gradient $dR(\theta)/d\theta$ around the dip a
small change $\Delta k_{\rm sp}$ in SP wave vector results in the
shift in the resonant angle $\Delta\Theta$ (or optical wavelength
$\Delta\lambda$) inducing a large change $\Delta R$ in
reflectivity - a physical principle behind the surface plasmon
resonance (SPR) sensing \cite{Homola99SAB54,Piliarik72OC6}.

The metal layer must be relatively thin (typically $\sim$50nm) in
order to allow for efficient coupling of incident light from the
glass-metal to the metal-air interface
\cite{Fukui79PSS91,Sarid81PRL47}. Incident light must be
p-polarized  for optimum coupling to SP: indeed, the electric
fields of s-polarized beam and SP are perpendicular making their
coupling impossible. Due to phase-matched excitation mechanism in
Kretschmann geometry SP intensity grows upon propagation along the
surface and can get up to 300 times larger as compared to the
incident light intensity, depending on energy dissipation in the
metal \cite{Weber81OL6}. In a realistic situation the enhancement
factor is lower due to other effects reducing the surface plasmon
propagation length such as, for example, surface roughness
\cite{Nagpal09Science325}. This SP intensity enhancement plays a
major role in the non-linear optics with surface plasmons
\cite{Grosse2012PRL108} and in the generation of high-energy
electrons with ultrashort SP pulses.

Another way of SP excitation is based on light scattering on metal
nano-structures. The plasmonic double-slit experiment
\cite{Schouten05PRL94_053901} triggered the development of
quantitative surface plasmon interferometry
\cite{Gay06NPh2_262,Temnov07OL32,Pacifici07NPhot1,Temnov10NPhoton4}.
Figure 6c illustrates SP interferometry with a slit-groove
arrangement in gold film. A collimated laser beam (either
continuous or pulsed) illuminates the entire area of the
microinterferometer and leads to most efficient SP excitation at a
wider groove. SPs propagate in the direction perpendicular to
groove axes towards the slit, where they are scattered back into
free-space radiation and interfere with directly transmitted
light. In case of a {\it tilted} groove
\cite{Temnov10NPhoton4,SPInterferometry09OE} the slit-groove
distance $D(x)$ varies along the slit axes $x$ resulting into the
periodic modulation of light intensity transmitted through the
slit. Any modulation of SP wave vector $k_{\rm sp}$ or propagation
distance $L_{\rm sp}$ induce the phase shift or change in contrast
of this plasmonic interference pattern, respectively, see example
in Fig.~4c.

%\bibliography{MagnetoPlasmonicsScienceBib}

%\newpage
%\textbf{Acknowledgements}\\ \\

%\textbf{Author contributions}\\ All authors have contributed to
%this paper and agree to its contents.\\

%\textbf{Competing interests}\\ The authors declare that they have
%no competing financial interests.\\

 \textbf{Correspondence}\\Correspondence and requests for materials should be addressed to
V.T.~(email: vasily.temnov@univ-lemans.fr).

%%
%% TABLES
%%
%% If there are any tables, put them here.
%%

\end{document}